\documentclass[twocolumn,prl,superscriptaddress,showpacs,preprintnumbers,amsmath,amssymb]{revtex4-1}

\usepackage[pdftex]{graphicx}
\usepackage{epstopdf}
\usepackage{graphicx}% Include figure files
\usepackage{dcolumn}% Align table columns on decimal point
\usepackage{bm}% bold mathx
\usepackage{amssymb}%symbols of the figures
\usepackage{wasysym}%other symbols of the figures
\usepackage{amsmath} 
\usepackage{amsthm}
\usepackage{bibtopic}
\usepackage{sidecap}
\usepackage{hyperref}
\usepackage[rflt]{floatflt}

\begin{document}

\def\Ef{$E_{\rm F}$}
\def\Eb{$E_{\rm B}$}
\def\Efmath{E_{\rm F}}
\def\Ed{$E_{\rm D}$}
\def\Tc{$T_{\rm C}$}
\def\kpara{k$_\parallel$}
\def\kparamath{{\bf k}_\parallel}
\def\kperp{{k}$_\perp$}
\def\Gbar{$\overline{\Gamma}$}
\def\Kbar{$\overline{K}$}
\def\Mbar{$\overline{M}$}
\def\BiTe{Bi$_2$Te$_3$}
\def\BiSe{Bi$_2$Se$_3$}
\def\SbTe{Sb$_2$Te$_3$}
\def\Ed{$E_{\rm D}$}
\def\invA{\AA$^{-1}$}
\def\Ef{$E_{\rm F}$}
\def\Tc{$T_{\rm C}$}
\def\kpara{{k}$_\parallel$}
\def\kperp{{k}$_\perp$}
\def\dirGX{$\overline{\rm \Gamma}-\overline{\rm X}$}
\def\dirGY{$\overline{\rm \Gamma}-\overline{\rm Y}$}
\def\pntG{$\overline{\rm \Gamma}$}
\def\pntX{$\overline{\rm X}$}
\def\invA{\AA$^{-1}$}
\def\G{$\Gamma$} 
\def\Z{$Z$}

\title{Impact of ultrafast transport on the high-energy states of\\ a photoexcited topological insulator}
\author{F. Freyse}
\affiliation{Helmholtz-Zentrum Berlin f\"ur Materialien und Energie, Albert-Einstein-Str. 15, 12489 Berlin, Germany}
\author{M. Battiato}
\affiliation{Nanyang Technological University, Nanyang Link 21, 637371, Singapore}
\author{L. V. Yashina}
\affiliation{Department of Chemistry, Moscow State University, Leninskie Gory 1/3, 119991, Moscow, Russia}
\author{J. S\'anchez-Barriga}
\email[Corresponding author. E-mail address: ] {jaime.sanchez-barriga@helmholtz-berlin.de.}
\affiliation{Helmholtz-Zentrum Berlin f\"ur Materialien und Energie, Albert-Einstein-Str. 15, 12489 Berlin, Germany}

\begin{abstract}

Ultrafast dynamics in three-dimensional topological insulators (TIs) opens new routes for increasing the speed of information transport up to frequencies thousand times faster than in modern electronics. However, up to date, disentangling the exact contributions from bulk and surface transport to the subpicosecond dynamics of these materials remains a difficult challenge. Here, using time- and angle-resolved photoemission, we demonstrate that driving a TI from the bulk-conducting into the bulk-insulating transport regime allows to selectively switch on and off the emergent channels of ultrafast transport between the surface and the bulk. We thus establish that ultrafast transport is one of the main driving mechanisms responsible for the decay of excited electrons in prototypical TIs following laser excitation. We further show how ultrafast transport strongly affects the thermalization and scattering dynamics of the excited states up to high energies above the Fermi level. In particular, we observe how inhibiting the transport channels leads to a thermalization bottleneck that substantially slows down electron-hole recombination via electron-electron scatterings. Our results pave the way for exploiting ultrafast transport to control thermalization time scales in TI-based optoelectronic applications, and expand the capabilities of TIs as intrinsic solar cells.

\end{abstract}
 
%\pacs{}  

\maketitle

\section {I. Introduction}

Topological insulators (TIs) are a unique state of matter \cite{KanePRL2005, KoenigScience2007, Fu-PRL-2007, Hasan-RMP-2010, Moore-Nature-2010} that allows generation and manipulation of dissipationless pure spin currents and spin-polarized electrical currents on ultrafast time scales \cite{Gedik-2011-PRL-Kerr, Gedik-2012-NatNanotech-photocurrents, Kastl-NatComm-2015, Sanchez-Barriga-PRB-2016, Boschini-SciRep-2016}. Their control is critically important for the application of TI materials in ultrafast spintronics and optoelectronics \cite{Das-Sarma-RMP-2004,Pesin-NatMat-2012,Carva-NatPhys-2014}, especially for the realization of low-power consumption devices operating at much higher speed than devices currently available for applications in modern electronics.

A promising route for the excitation of ultrafast currents on TI surfaces is the use of femtosecond (fs) laser pulses \cite{Gedik-2012-NatNanotech-photocurrents, Kastl-NatComm-2015, Sobota-PRL-2012-Bulk-Reservoir, Gedik-2013-Science-Floquet, Kuroda-PRL-2016}. Due to strong spin-orbit coupling causing a band inversion in the bulk, such currents can be driven under nonequilibrium conditions through Dirac-cone-like topological surface states (TSSs) that are protected by time-reversal symmetry \cite{Hsieh-Nature-2009,Roushan, LeePRB09, Valla, Sanchez-Barriga-PRB-warp, Sanchez-Barriga-NatComm-2016} and characterized by a helical spin texture where electron spins are locked to their linear momentum \cite{Hsieh-Nature-2009, Hsieh-Science-2009, Jozwiak-PRB-2011, Pan-PRB-2013, Sanchez-Barriga-PRX-2014, Zhu-PRL-2014}. The topological protection renders the generated currents robust against non-magnetic impurities, and thus their properties are fundamentally distinct than when induced, for instance, in conventional semiconductors or topologically trivial matter \cite{Battiato1, Battiato2}. 

The generation of ultrafast currents in TIs requires however the excitation of a nonequilibrium population of hot electrons above the Fermi level on ultrafast time scales \cite{Gedik-2011-PRL-Kerr, Gedik-2012-NatNanotech-photocurrents, Braun-NatComm-2016}. Equally important is to realize a versatile control of the relaxation time scales of the generated currents, especially if the information encoded by the spin and charge degrees of freedom is required to travel over macroscopic distances \cite{Das-Sarma-RMP-2004}. Therefore, a lot of attention is being devoted to the fundamental understanding of the underlying mechanisms responsible for the ultrafast relaxation of hot carriers on TI surfaces following laser excitation.

Time-resolved angle-resolved photoemission (tr-ARPES) is a very powerful tool for such a purpose, as it allows to probe the ultrafast dynamics of photoexcited carriers and elementary scattering process directly in the electronic band structure with high temporal, energy and momentum resolution \cite{Bovensiepen-2012, Lanzara-Perspective-2016}. By means of tr-ARPES, bulk-assisted electron-phonon scattering has been identified as one of the relevant mechanisms responsible for the decay process of hot electrons across the linear energy-momentum dispersion of TSSs \cite{Gedik-2011-PRL-Kerr, Sobota-PRL-2012-Bulk-Reservoir, Hajlaoui-2012-Nanolett-bulk, Gedik-2012-PRL-phonons, Luo-NanoLett-2013-increased-e-ph}. The influence of collective modes such as phonons in the dynamics of TIs has also been evidenced by the emergence in tr-ARPES of coherent-phonon oscillations both at the surface and in the bulk \cite{Sobota-PRL-2014-Oscillations, Golias-PRB-2016-Oscillations}. By combining tr-ARPES with spin resolution, it has been shown that the characteristic time scale for hot electron relaxation in TSSs is in addition influenced by electron-electron scattering processes that ultimately depend on the complex alternating spin texture of excited states above the Fermi level \cite{Cacho-Surface-resonances-PRL-2015, Jozwiak-NatComm-2016, Sanchez-Barriga-PRB2-2017}. 

In particular, the tunability of the relaxation time scales of TSSs observed by tr-ARPES has been primarily realized by the action of band bending on the interplay between surface and bulk carrier dynamics \cite{Hajlaoui-NatComm-2014-e-h-asym}, as for instance via bulk doping \cite{Ando-JPSoc-2013,JZhang-NatComm-2011}. This effect allowed the creation of a long-lived surface photovoltage \cite{Neupane-PRL-2015,Sanchez-Barriga-APL-2017}, and it is believed to be responsible for an electron-hole asymmetry that slows down the relaxation dynamics \cite{Hajlaoui-NatComm-2014-e-h-asym}. Changes in the decay rates of excited carriers in TIs have been also observed as a function of temperature, momentum and energy. The counter-intuitive temperature-dependence of the decay rates observed previously in a prototypical TI has been primarily attributed to electron-phonon scattering processes \cite{Sobota-JSRP-2014}. Interestingly, the fact that the population relaxation in a TI becomes slower for increasing temperature was explained on the basis of phonon emission and absorption processes from the excited hot-electron population \cite{Sobota-JSRP-2014}. However, despite all the recent important advances in our understanding of the role of these elementary scattering processes in TIs, the impact of ultrafast transport in a photoexcited TI especially up to high energies above the Fermi level has remained elusive so far.

More generally, the impact of out-of-equilibrium transport on the ultrafast dynamics of condensed-matter systems such as semiconductors or magnetic materials has been put forward only recently \cite{Battiato3, Battiato4, Battiato5}. The concept of ultrafast transport is fundamentally different than conventional carrier transport. The main difference is not only that transport properties of states well above the Fermi energy have to be accounted for, but also the ultrashort time scales involved. During such short times external electric fields have only a secondary effect on transport, which is instead mainly driven by superdiffusion of highly-excited electrons \cite{Battiato3, Battiato4, Battiato5}. In the theoretical work where this picture was put forward \cite{Battiato3}, it was shown how laser-excited electrons undergo superdiffusion which is strongly dependent on their energy and spin. This effect was shown to be the key to ultrafast demagnetization \cite{Beaurepaire-1995, Malinowski-2008}, and to play an important role in high-efficiency THz emission \cite{Kampfrath-NatNano-2013, Seifert-NatNano-2013} and in the transmission of charge and spin on ultrafast time scales \cite{Battiato1, Battiato2, Malinowski-2008, Melnikov-PRL-2011, Rudolf-NatComm-2011, Eschenlohr-NatMat-2013, Graves-NatMat-2013, Vodungbo-NatComm-2012}.

Therefore, in the present work we focus our attention on the role of ultrafast transport in the relaxation dynamics of TIs following fs-laser excitation. By means of tr-ARPES, we establish that ultrafast transport processes between the surface and the bulk is one of the most important channels for the relaxation of carriers in prototypical TIs. We demonstrate how ultrafast transport can be inhibited by driving a TI from the bulk-conducting into the bulk-insulating regime, and especially how it strongly affects the thermalization and scattering dynamics of carriers. We further show how ultrafast transport impacts the dynamics of the excited states up to high energies above the Fermi level, and unravel the existence of a thermalization bottleneck for electron-hole recombination via elementary scattering processes. Finally, we discuss important implications of our findings in the context of TI-based applications.

\section {II. Experimental details}
 
We performed synchrotron- and laser-based tr-ARPES experiments to investigate both the electronic structure in equilibrium and the ultrafast dynamics of excited states in TIs at room temperature, respectively. The synchrotron-based ARPES measurements were carried out using linearly-polarized undulator radiation at the UE112-PGM2 beamline of the synchrotron source BESSY-II in Berlin. Photoelectrons were detected with a Scienta R8000 electron analyzer and the base pressure of the experimental setup was better than $1\cdot10^{-10}$ mbar. Laser-based tr-ARPES experiments were performed using linearly-polarized 1.5 eV pump and 6 eV probe fs-laser pulses focused on the sample, and by varying the pump-probe time delay $\Delta$t with an optical delay stage. We used a home-made Ti:Sapphire fs oscillator coupled to an ultrafast amplifier laser system (RegA, Coherent) operating at 150 kHz repetition rate. The time resolution of the experiment was $\sim$200 fs, and the pump fluence $\sim$100 $\mu$J/cm$^{2}$. The angular and energy resolutions of the experiments were 0.1$^{\circ}$ and 20 meV, respectively.

Experiments were performed on Bi$_2$Te$_3$ and Bi$_{0.9}$Sb$_{1.1}$Te$_3$ bulk single crystals cleaved {\it in situ} and grown by the Bridgman method \cite{Shtanov-JCG-2009}. The crystals were grown using different compositions from the melt. The growth temperatures were in the range of 590--600$^{\circ}$C, and the temperature gradient during growth was 5$^{\circ}$C/cm. The structure and sample composition were characterized by various spectroscopic and diffraction methods, as detailed elsewhere \cite{Volykhov-submitted-2017}. The high crystal quality of the obtained (111) surfaces was verified by low-energy electron diffraction as well as by the presence of sharp features in the ARPES dispersions.

\begin{figure*}
\centering
\includegraphics [width=0.75\textwidth]{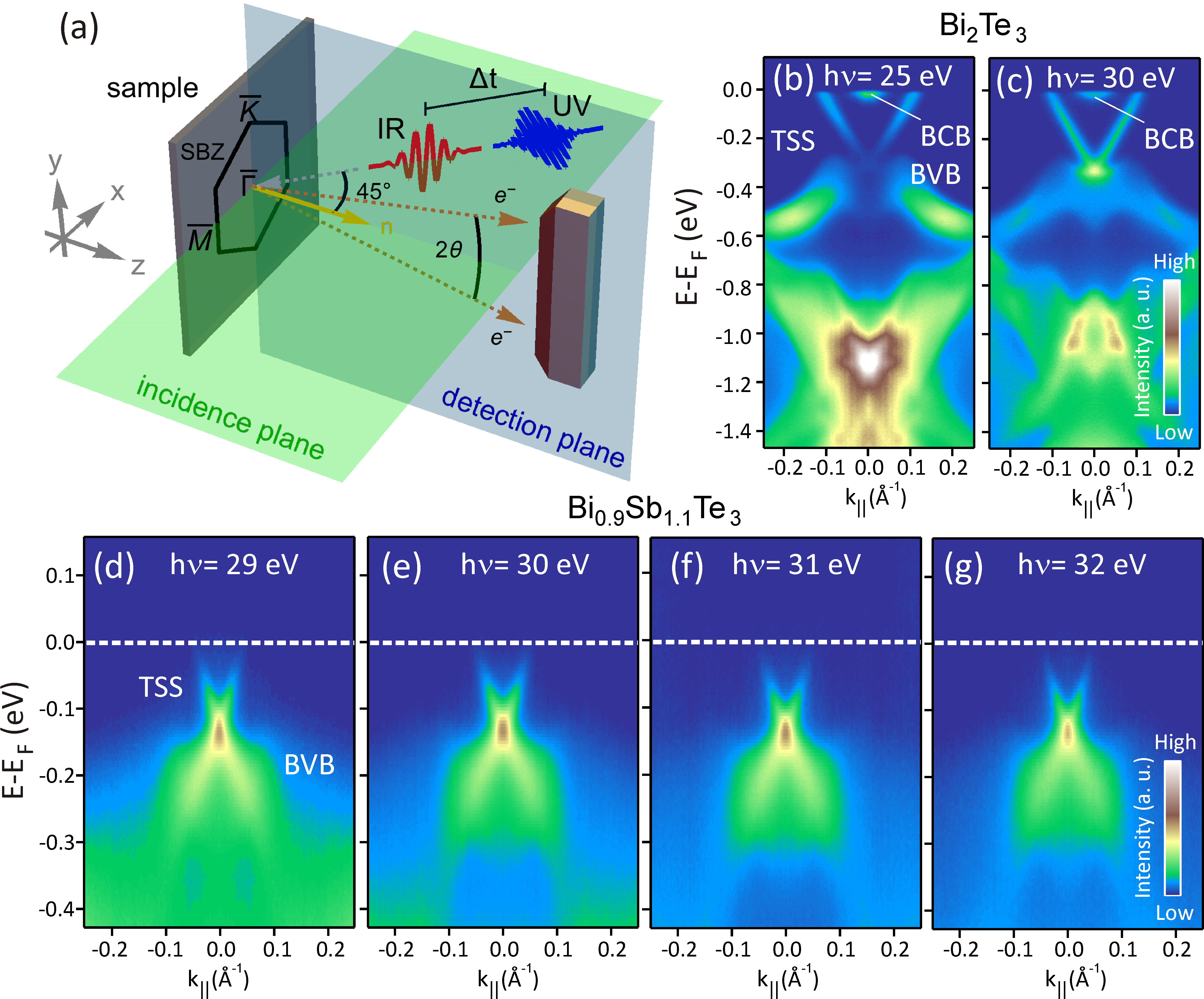}
\caption{(Color online) (a) Geometry of the synchrotron- and laser-based photoemission experiments. The light incidence and electron emission planes are indicated in green (light) and blue (dark) colors, respectively. The detection plane is oriented along the \Gbar\Kbar\ direction of the SBZ, and the light impinges the sample under an angle of $\phi=45^{\circ}$ with respect to the surface normal. The tr-ARPES measurements are taken by varying the time delay $\Delta$t between infrared (1.5 eV) pump and ultraviolet (6 eV) probe pulses of vertical and horizontal linear polarization, respectively. (b)-(g) ARPES dispersions of (b),(c) Bi$_2$Te$_3$ and (d)-(g) Bi$_{0.9}$Sb$_{1.1}$Te$_3$ taken under equilibrium conditions using linearly-polarized synchrotron light of different photon energies.}
\label{equil}
\end{figure*}

Measurements were performed using the sample geometry shown in Fig.~\ref{equil}(a), where either the synchrotron light or the fs-laser pulses were incident on the sample under an angle of $\phi=45^{\circ}$ with respect to the surface normal. Photoelectrons were collected up to acceptance angles of $\pm$15$^{\circ}$ using an analyzer entrance slit size of 0.2 mm $\times$ 25 mm. The long axis of the entrance slit, which defines the electron detection plane, was parallel to the \Gbar\Kbar\ direction of the surface Brillouin zone (SBZ).

\section {III. Results and Discussion} 

To investigate the electronic properties of Bi$_2$Te$_3$ and Bi$_{0.9}$Sb$_{1.1}$Te$_3$ in equilibrium, we first performed synchrotron-based ARPES experiments as a function of photon energy [see Figs.~\ref{equil}(b)-(g)]. Quantization effects and a Rashba splitting of bulk states indicating a strong band bending at the surface were not observed \cite{King-PRL-2011}. Changing the photon energy allowed us to select the component of the electron wave vector perpendicular to the surface \kperp \cite{Hufner-2007}, and to directly visualize whether the characteristic \kperp\ dispersion of bulk states leads to contributions from the bulk-conduction band (BCB) or bulk-valence band (BVB) at the Fermi level. This type of measurements are important to understand the ultrafast response of excited states to laser excitation, as one might expect that the presence of bulk states at the Fermi level can substantially increase the phase space available for the relaxation of excited carriers. Therefore, the possibility of combining synchrotron-based ARPES and laser-based tr-ARPES in one experiment offers clear advantages with respect photoemission experiments solely based on ultrafast laser sources, where usually the photon energy is either fixed or cannot be systematically varied in a wide range of desired values. We also emphasize that because our experiments are surface sensitive (with probing depths between 1 nm at 50 eV and $<$10 nm at 6 eV) \cite{Hufner-2007}, in the present work we are only addressing the equilibrium band structure, its time evolution, and dynamical effects within the near surface region.
\begin{figure*}
\centering
\includegraphics [width=0.85\textwidth]{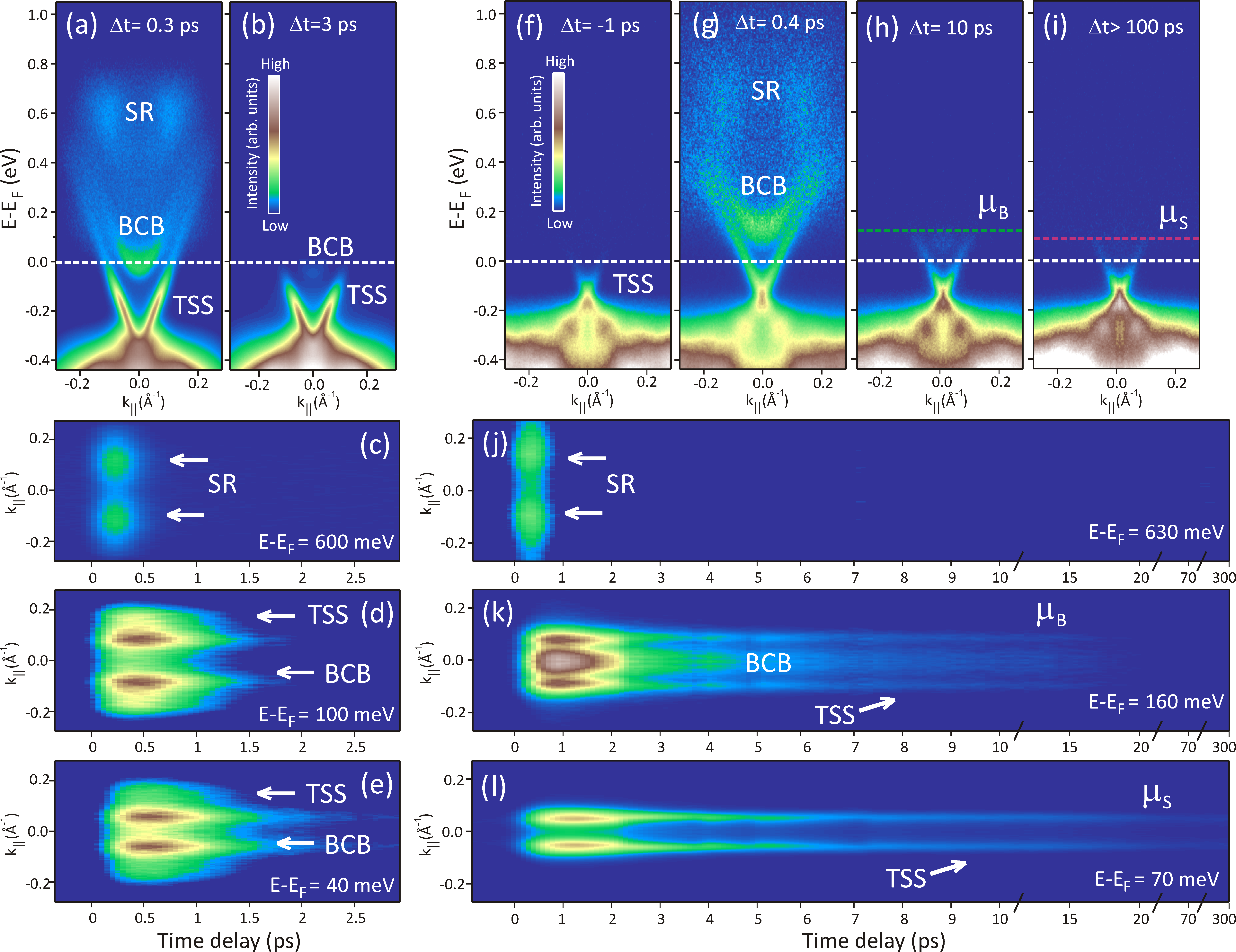}
\caption{(Color online) Ultrafast dynamics of excited states following optical excitation by fs-laser pulses in (a)-(e) Bi$_2$Te$_3$ and (d)-(l) Bi$_{0.9}$Sb$_{1.1}$Te$_3$. High-resolution tr-ARPES spectra at selected pump-probe delays are shown in (a),(b) and (f)-(i), respectively. Correspondingly, the time evolution of the momentum-resolved intensities at various energies above the Fermi level is shown in (c)-(e) and (j)-(l). Intensity contributions from a topologically trivial surface resonance, the bulk-conduction band and topological surface state are denoted as SR, BCB and TSS, respectively. The metastable long-lived populations associated to the bulk and the surface are labelled $\mu_B$ and $\mu_S$ in (h),(i),(k) and (l).}
\label{dynam}
\end{figure*} 

Figures~\ref{equil}(b) and \ref{equil}(c) display high-resolution ARPES dispersions of the TSS, BCB and BVB states of Bi$_2$Te$_3$ measured in equilibrium with 25 and 30 eV photons, respectively. We observe a gapless Dirac cone representing the TSS together with clear intensity contributions from the BCB at the Fermi level at both photon energies. The Dirac point of the TSS is located at an energy of E-E$_{F}\sim$ -0.3 eV, indicating that the sample is intrinsically $n$-doped. The energy position of the BCB, which reaches a minimum at E-E$_{F}\sim$-42 meV in Fig.~\ref{equil}(c), proves that the sample is in the bulk-conducting transport regime with the Fermi level outside the bulk band gap, in agreement with previous observations \cite{Ando-JPSoc-2013,JZhang-NatComm-2011,Ren-PRB-2011}. It is also seen that the Dirac point is buried in the BVB, which reaches a maximum energy of E-E$_{F}\sim$-240 meV at off-normal wave vectors. In consequence, the Dirac point will never be accessible by conventional transport experiments even if it would be shifted up to the Fermi level by using e.g., an external gate voltage. 
\begin{figure*}
\centering
\includegraphics [width=0.725\textwidth]{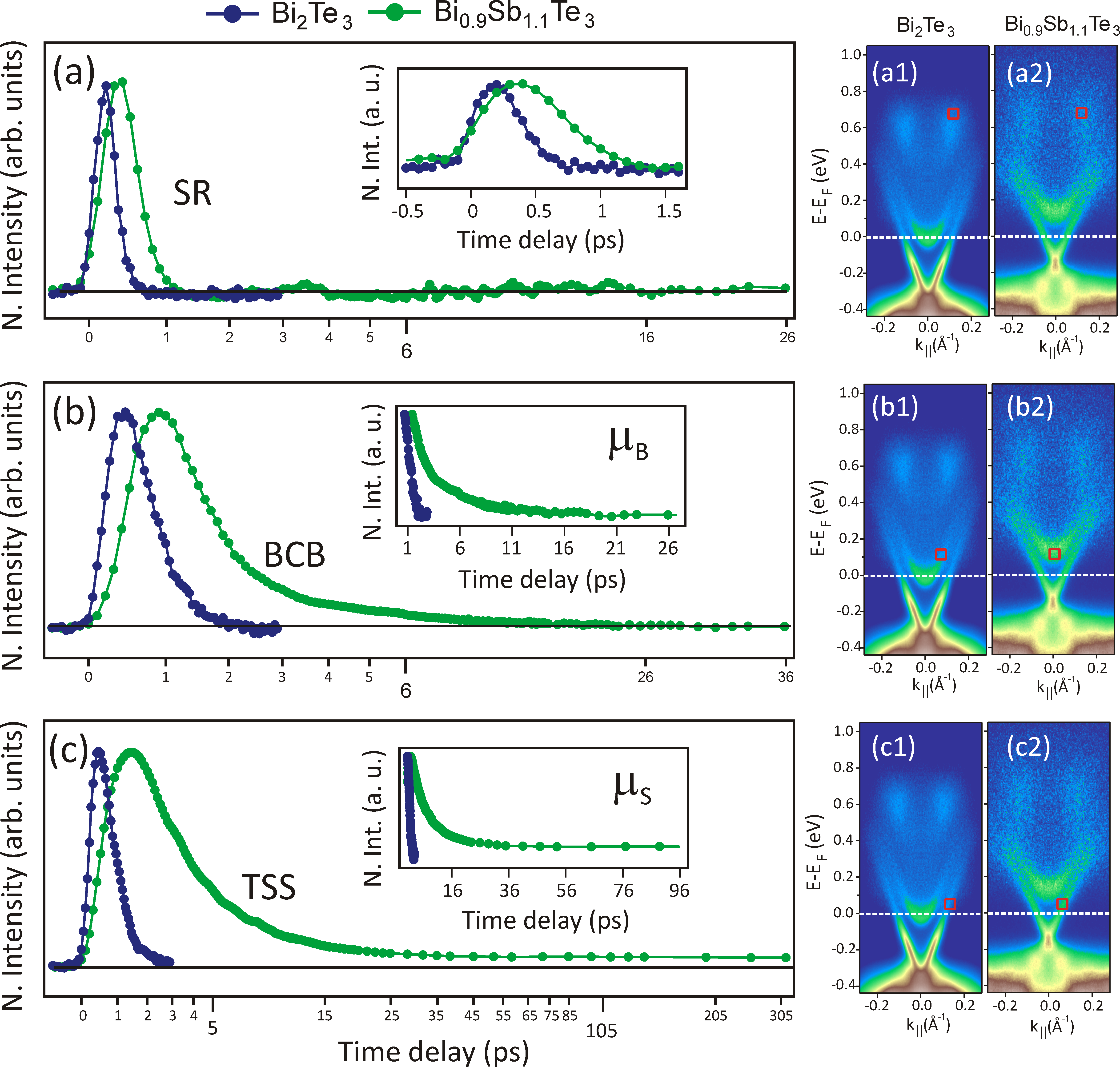}
\caption{(Color online) (a)-(c) Normalized tr-ARPES intensities from different bands as a function of pump-probe delay for Bi$_2$Te$_3$ (blue circles) and Bi$_{0.9}$Sb$_{1.1}$Te$_3$ (green circles). For each band, the intensities are integrated within small energy-momentum windows that are at the same energy for the bulk-conducting and bulk-insulating case [see panels (a1)-(c2)]. Changes in the dynamics of (a) surface resonance (SR) states, (b) bulk-conduction band (BCB) and (c) topological surface state (TSS) are also emphasized in the corresponding insets. The labels $\mu_B$ and $\mu_S$ in (b) and (c) denote the metastable long-lived populations associated to the bulk and the surface, respectively.}
\label{relax}
\end{figure*} 

This particular situation appears to be completely different for Bi$_{0.9}$Sb$_{1.1}$Te$_3$ [see Figs.~\ref{equil}(d)-\ref{equil}(g)], where both the Dirac point and the Fermi level lie within the bulk band gap as proven by the photon-energy dependence of the ARPES dispersions. Besides the fact that the TSS has no $k_z$ dependence due to its two-dimensional nature, the Dirac point has moved upwards by about $\sim$0.1 eV and the contributions from BCB states observed for Bi$_2$Te$_3$ at the Fermi level entirely disappear. The upward shift with respect to Bi$_2$Te$_3$ is also accompanied by an increase in the group velocity of the TSS from 0.32 to 0.44 nm/fs. We do also note that due to the large lattice constant of Bi$_{0.9}$Sb$_{1.1}$Te$_3$ along the $z$ direction \cite{Volykhov-submitted-2017}, in Figs.~\ref{equil}(d)-\ref{equil}(g) we practically cross the complete bulk Brillouin zone without observable contributions from bulk states at the Fermi level. These findings strongly indicate that Bi$_{0.9}$Sb$_{1.1}$Te$_3$ is within the bulk-insulating transport regime, and pinpoint the role of hole-type bulk carriers induced by Sb--Te antisite defects in the modification of the electronic band structure \cite{JZhang-NatComm-2011}. The fact that the sample becomes truly bulk-insulating is also consistent with observations of the surface quantum Hall effect \cite{Yoshimi-NatComm-2015} and the ambipolar field effect \cite{Kong-NatNanotech-2011} in samples of the same critical composition. 

If we now compare the dynamics of excited states in Bi$_{2}$Te$_{3}$ [Figs.~\ref{dynam}(a)-\ref{dynam}(e)] and Bi$_{0.9}$Sb$_{1.1}$Te$_3$ [Figs.~\ref{dynam}(a)-\ref{dynam}(l)] following laser excitation, we find remarkable changes in the decay times up to high energies above the Fermi level when compared to the energy of the exciting pump. While the overall dynamics in the case of Bi$_2$Te$_3$ is as fast as $\sim$ 3 ps [Figs.~\ref{dynam}(a) and \ref{dynam}(b)], excited electrons within the TSS of Bi$_{0.9}$Sb$_{1.1}$Te$_3$ persist on a much longer time scale which exceeds the measured time window [Figs.~\ref{dynam}(a)-\ref{dynam}(i)]. The effect is also seen directly in the time dependence of the momentum-resolved intensities extracted at different energies above the Fermi level [Figs.~\ref{dynam}(c)-\ref{dynam}(e) and \ref{dynam}(j)-\ref{dynam}(l)]. In addition, we clearly observe signatures of two metastable long-lived populations associated to the bulk and the surface [labelled $\mu_B$ in Figs.~\ref{dynam}(h) and \ref{dynam}(k) and $\mu_S$ in Figs.~\ref{dynam}(i) and \ref{dynam}(l), respectively]. But what is most surprising is the fact that the relaxation dynamics of the states located at energies around $\sim$0.6--0.8 eV above the Fermi level (labelled SR in Figs.~\ref{dynam}(a), \ref{dynam}(c), \ref{dynam}(g) and \ref{dynam}(j)] considerably slows down when the sample becomes truly bulk insulating [compare, e.g., Figs.\ref{dynam}(c) and \ref{dynam}(j)]. This effect, which was  previously unrealized in investigations of the ultrafast dynamics of TIs, recalls the need of a much more refined interpretation of the underlying mechanisms responsible for the observed electron dynamics.
\begin{SCfigure*}
\centering
\includegraphics [width=0.7\textwidth]{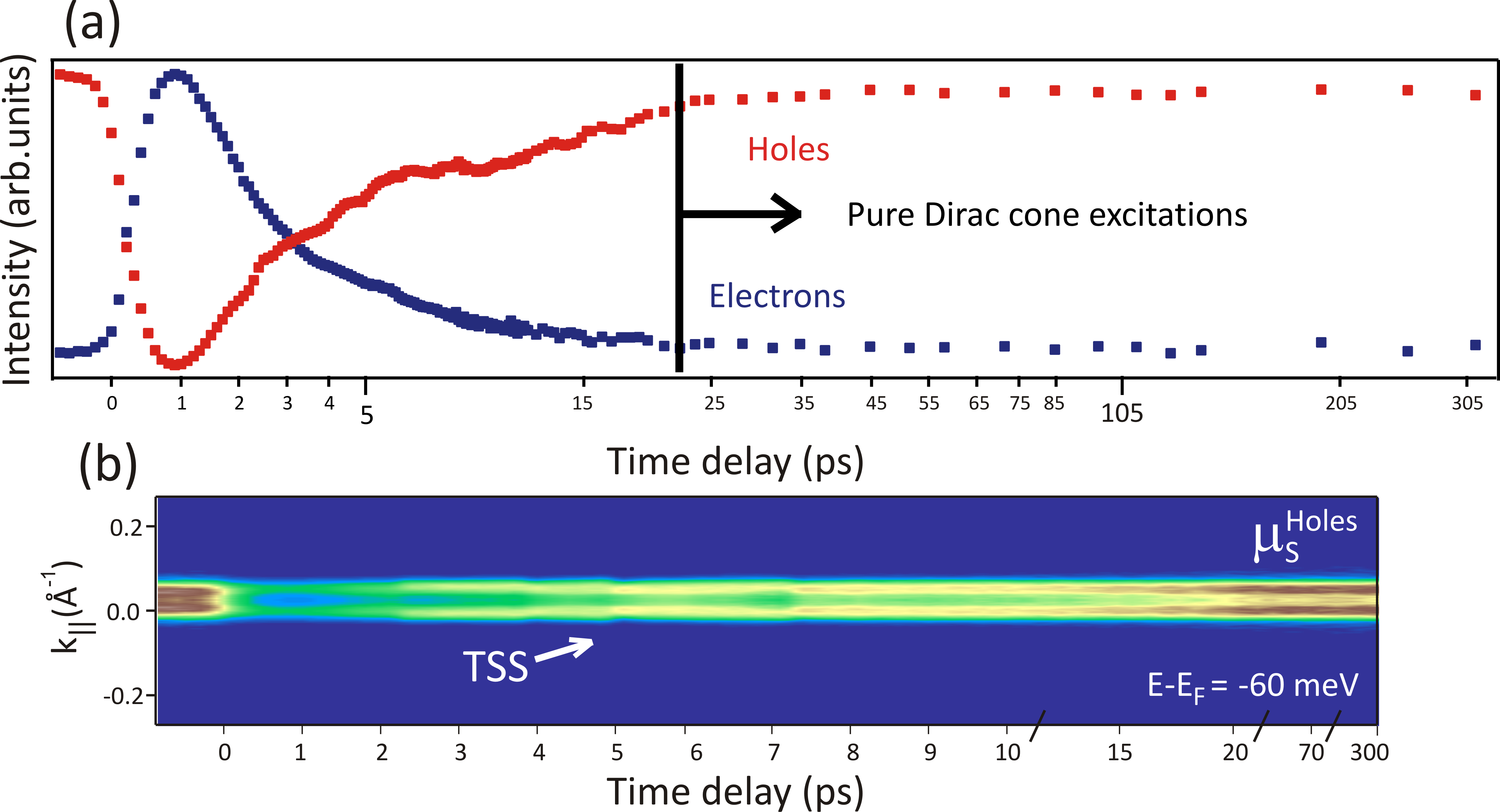}
\caption{(Color online) (a) Comparison between tr-ARPES intensities extracted within the TSS at an energy of $\sim$60 meV above and below the Fermi level in Bi$_{0.9}$Sb$_{1.1}$Te$_3$, depicting the dynamics of electrons (blue) and holes (red). Once the metastable population in the bulk-conduction band completely relaxes (indicated by a vertical thick line), electrons and holes within the TSS are the only contribution to the measured signal. (b) Time evolution of the momentum-resolved intensity of the TSS bands below the Fermi level corresponding to the hole dynamics shown in (a).}
\label{holexc}
\end{SCfigure*} 

We have previously pointed out the crucial role of electron-electron scattering processes in the relaxation dynamics of the SR states in Bi$_2$Te$_3$ \cite{Sanchez-Barriga-PRB2-2017}. These states form a surface resonance that disperses inside a projected bulk band gap located at higher energy than the main gap of the volume \cite{Sanchez-Barriga-PRB2-2017}. However, the fact that their dynamics slows down in the bulk-insulating transport regime as seen in Fig.~\ref{dynam} rises the question of which additional mechanism is fundamentally contributing to the relaxation of carriers. We emphasize that the existence of BCB states at the Fermi level in the bulk-conducting transport regime is difficult to reconcile with an accelerated electron dynamics of SR states. The reason is that electron-electron scatterings are not efficient in redistributing energy. In particular, provided the large density of states of the BCB at the Fermi level in Bi$_2$Te$_3$, the most likely outcome of a direct relaxation channel into the BCB states will be the excitation of a second electron into the SR states under the conservation of energy and momentum. Conversely, this process would be responsible for a slower relaxation dynamics of SR states in the bulk-conducting transport regime, differently from our observations. 

Moreover, the changes in the decay times observed when comparing Figs.~\ref{dynam}(c)-\ref{dynam}(e) to Figs.~\ref{dynam}(i)-\ref{dynam}(l), respectively, are also difficult to reconcile with the expected energy dependence of the decay times, according to which states at higher energy should have faster relaxation. In fact, the momentum-resolved intensities displayed in Figs.~\ref{dynam}(i)-\ref{dynam}(l) exhibit much longer relaxation times but are extracted at higher energies than the corresponding momentum-resolved intensities shown in Figs.~\ref{dynam}(c)-\ref{dynam}(e). One might suggest that the overall shift of the electronic band structure or even the linearity of the bands is responsible for this behavior. However, that explanation would be completely at odds with the extremely similar relaxation times of the order of few ps observed for Bi$_2$Te$_3$ \cite{Sanchez-Barriga-PRB2-2017} and Sb$_2$Te$_3$ \cite{Sanchez-Barriga-PRB-2016}, despite the strong upward shift in the energy of the Dirac point and the change in the linearity of the surface bands when going from the former to the latter \cite{JZhang-NatComm-2011}.

The changes in the decay times can also be observed in the tr-ARPES intensities
integrated at the same energy for the bulk-conducting and bulk-insulating case, and within small energy-momentum windows among different states, as shown in Fig.~\ref{relax}. For the SR states [Fig.~\ref{relax}(a)], it is remarkable that both the rise and decay times become slower when the sample is truly bulk-insulating [see also inset of Fig.~\ref{relax}(a)]. The slower rise time of SR states strongly indicates a delayed thermalization in the bulk-insulating case. Otherwise, the relaxation of SR states proceeds according to a time constant of $\tau$=338$\pm$22 fs which is more than twice as large than $\tau$=163$\pm$15 fs for the bulk-conducting case. 

We find a similar situation for states within the bulk-conduction band [Fig.~\ref{relax}(b)]. However, their relaxation in the bulk-insulating regime proceeds according to two time constants $\tau_1$=0.96$\pm$0.02 ps and $\tau_2$=7.5$\pm$0.8 ps [as also seen in the inset of Fig.~\ref{relax}(b)]. This behavior is in contrast to the fact that the relaxation of bulk states in the  bulk-conducting case proceeds according to one time constant ($\tau$=0.57$\pm$0.03 ps). Likewise, the relaxation of the TSS [see Fig.(c) and inset] in the bulk-insulating (bulk-conducting) regime proceeds according to $\tau_1$=2.85$\pm$0.3 ps and $\tau_2$=690$\pm$20 ps ($\tau_2$=0.63$\pm$0.03 ps), meaning that the effect is substantially more pronounced particularly considering the prominent increase of $\tau_2$ which represents a metastable electron population on the surface ($\mu_S$). In this respect, it is also remarkable that holes within the TSS right below the Fermi level exhibit similar dynamics [see Fig.~\ref{holexc}(a)]. At long time delays, as also seen Fig.~\ref{holexc}(b), there is a metastable population of holes in the vicinity of the Fermi level ($\mu^{Holes}_S$). Strictly speaking, once the metastable population $\mu_B$ completely relaxes [indicated by a vertical thick line in Fig.~\ref{holexc}(a)], long-lived electrons and holes localized within the TSS bands are the only contribution to the measured signal. However, it should be pointed out that this effect is fundamentally different than the previously proposed topological exciton condensation in TI films \cite{Seradjeh-PRL-2011}. 
\begin{figure*}
\centering
\includegraphics [width=0.85\textwidth]{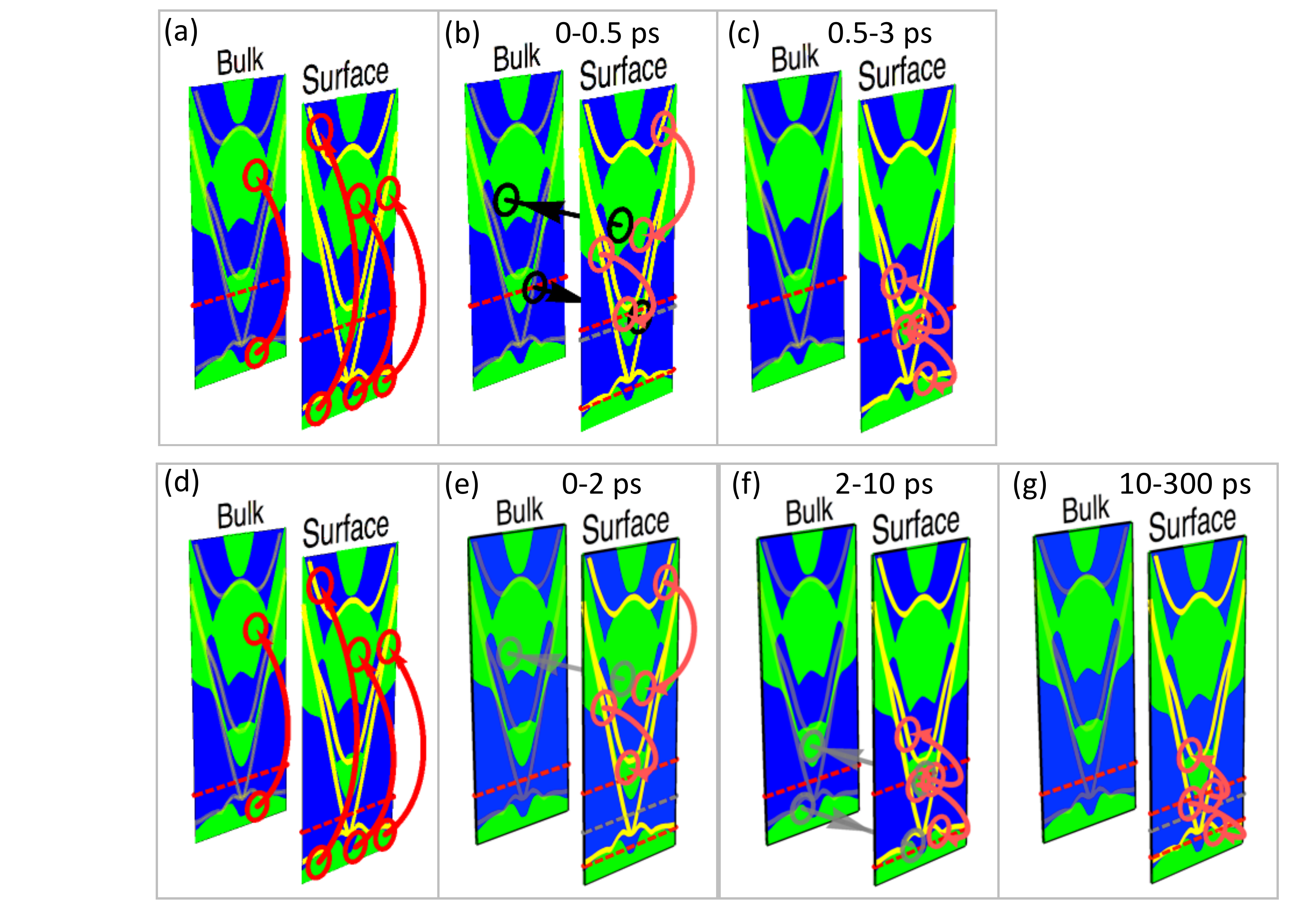}
\caption{(Color online) Schematics of the different thermalization stages. In each panel, the surface density of states is shown on the right, while the surface-projected bulk density of states is shown on the left. The typical time scales of the studied processes as observed experimentally are given on the top.  The surface states are plotted as semitransparent lines on the bulk density of states as a guide to the eye, and are not to be intended as real states in the bulk. The chemical potentials and effective chemical potentials are pictured as broken red lines. The dashed gray lines depict the chemical potential at previous delay times and used as a guide to the eye. The top row illustrates the thermalization stages in Bi$_2$Te$_3$. (a) The laser excites electrons from below the Fermi energy to empty states above (for simplicity, all transitions are shown as originating from the top of the valence band; however, it is understood that the original state of the excited electrons has to be the appropriate one below the Fermi energy). (b) Two thermalization channels are active: Scattering [orange arrows] and transport [black arrows]. (c) The recombination of electrons and holes is the last stage of thermalization. The bottom row depicts the thermalization stages in Bi$_{0.9}$Sb$_{1.1}$Te$_3$. (d) Laser excitation. (e) Electron-electron and electron-phonon scatterings contribute to the thermalization, but the transport is quickly suppressed by the formation of charged regions. (f) The first stage of electron-hole recombination is facilitated by the fact that the scattered electron can still be within the bottom of the conduction band having high density of states. (g) The second stage is slowed down by the fact that the second electron involved in the scattering has to come from the Dirac cone, which has a low density of states.}
\label{tstages}
\end{figure*} 

The faster thermalization observed in the bulk-conducting case, as well as the substantially larger relaxation times observed in the bulk-insulating regime, pinpoints ultrafast transport as one of the most important processes underlying the observed electron dynamics. Note that if impact ionization \cite{Zhu-SciRep-2015} played an important role in the filling of the states, we would expect a slower thermalization dynamics in the bulk-conducting case, in contrast to our observations (see Supplemental Materials for details \cite{Supplement}). Also note that by thermalization we refer to the situation in which the whole electronic system is (approximately) described by a global Fermi-Dirac distribution. It is known that ultrafast transport is a very important mechanism that strongly contributes to the thermalization and relaxation of carriers in condensed-matter systems following fs-laser excitation. Although neglected for a long time, transport of strongly out-of-equilibrium carriers has recently become the target of intense study \cite{Battiato3, Melnikov-PRL-2011, Rudolf-NatComm-2011, Kampfrath-NatNano-2013, Eschenlohr-NatMat-2013, Graves-NatMat-2013, Seifert-NatNano-2013, Bergeard-PRL-2016}. It has been shown to produce a wide range of effects \cite{Battiato1, Battiato2, Battiato3, Melnikov-PRL-2011, Rudolf-NatComm-2011, Kampfrath-NatNano-2013, Eschenlohr-NatMat-2013, Graves-NatMat-2013}, and to have an important impact on the fs dynamics in several materials \cite{Seifert-NatNano-2013, Bergeard-PRL-2016}, especially when the system is characterized by inhomogeneities (as for instance in multilayers) or when, simply (as in most of the cases), the laser excitation is confined very close to the surface. Studying thermalization without including the effect of out-of-equilibrium transport is likely to lead to misleading interpretations. Therefore, in the following we will focus our attention on the role of ultrafast transport in the context of our present observations for TIs. 

Let us firstly discuss the impact of ultrafast transport on the dynamics of the bulk-conducting case (Bi$_2$Te$_3$) [see Figs.~\ref{tstages}(a)-\ref{tstages}(c)]. On the right and left-hand side of each panel we plot the surface and the surface-projected bulk density of states, respectively \cite{Sanchez-Barriga-PRB2-2017}. The laser excites electrons from the valence band to the conduction band as shown in Fig.~\ref{tstages}(a) both on the surface and in the bulk. The excited electrons and holes start thermalizing due to two main mechanisms. Excited electrons [see Fig.~\ref{tstages}(b)] can scatter either with phonons (high momentum, low-energy transitions, not shown in figure) or with other electrons (highlighted by two possible transitions in orange). Note that for simplicity, we do not show the transitions compatible with energy, momentum and spin selection rules associated with the second electron participating in the scattering and which is ousted from below the Fermi energy.

However, there is another important process that will contribute to the thermalization of the carriers: the transport towards the bulk. Electrons localized at the surface, but occupying bands that extend into the bulk, can diffuse towards the bulk [indicated by a black arrow going from surface to bulk in Fig.~\ref{tstages}(b)]. In fact, the displacement of an electron from the surface to the bulk will leave the surface positively and the bulk negatively charged. The generated electric field will drag one electron at the Fermi level from the bulk to the surface [denoted by a black arrow going from bulk to surface in Fig.~\ref{tstages}(b)], restoring the neutrality of the two regions. We emphasize that by only measuring the high-energy population at the surface, such process cannot be disentangled from thermalization due to scatterings. Moreover, the removal of electrons from the bottom of the conduction bands will further hasten the thermalization due to scatterings in the surface states, which do not have a direct transport channel towards the bulk. These two processes together lead to a very fast thermalization [Fig.~\ref{relax}(a)]. 

Holes will undergo similar dynamics (not explicitly shown in Fig.~\ref{tstages}), with the only difference that, given the fact that valence bands in Bi$_2$Te$_3$ are less dispersive than the conduction bands, the transport is expected to have a smaller impact on their thermalization. Note that however at this stage there is still trapping of electrons (holes) in the bottom of the conduction (top of the valence) band. This is represented in Fig.~\ref{tstages}(b) by the presence of two chemical potentials on the surface (the equilibrium chemical potential is shown in gray for reference). In the long run, for a full thermalization, electrons brought in the conduction band need to recombine with holes across the band gap. Electrons and holes will recombine on a slower time scale by transferring the excess energy to another electron around the Fermi energy as shown in Fig.~\ref{tstages}(c). The slower dynamics is due to the fact that scatterings on the Dirac cone are rare due to the low density of states, despite the fact that the ultrafast transport channels of electrons and holes from bulk to surface and vice versa are continuously active [Figs.~\ref{relax}(b) and \ref{relax}(c)].

Very interestingly, in the bulk-insulating transport regime (Bi$_{0.9}$Sb$_{1.1}$Te$_3$) [see Figs.~\ref{tstages}(d)-\ref{tstages}(g)], completely different dynamics emerges both in the bulk and on the surface. The laser excites similar transitions [Fig.~\ref{tstages}(d)] as for the bulk-conducting case, and excited electrons can thermalize due to scatterings or transport [Fig.~\ref{tstages}(e)]. However, in this case the transport [shown as a semitransparent gray arrow in Fig.~\ref{tstages}(e)] is very quickly suppressed due to the lack of carriers around the Fermi level in the bulk: the transferred charge will not be screened; it will instead accumulate, creating electric fields that quickly prevent further injection of charge carriers and therefore transport. This suppresses one of the most important channels of thermalization at the surface. The decay of high-energy population will as a consequence be importantly slower than in the bulk-conducting case, due to the absence of a very effective thermalization mechanism [Fig.~\ref{relax}(a)]. 

Similarly to what seen before, the electron-hole recombination [represented by the transient chemical potentials for electrons and holes in Fig.~\ref{tstages}(e)] is expected to take longer time. In this case, however, the existence of two effective potentials is extremely evident, at least for excited electrons. In consequence, electron-hole recombination [Fig.~\ref{tstages}(f)] takes place via two mechanisms, scattering and transport [gray arrows in Fig.~\ref{tstages}(f)]. The transport requires the simultaneous displacement of an electron (hole) in the conduction (valence) band from the surface to the bulk. This channel proceeds according to the slower carriers which in this case are the holes. Note that in Fig.~\ref{tstages}(g) the transport is completely blocked and only scattering remains as the main recombination mechanism. Therefore, electron-hole recombination will proceed in two time scales. We also note that the presence of the non-effective transport channels shown as gray arrows in Fig.~\ref{tstages}(f) explains why the decay of the population $\mu_B$ within the conduction band proceeds on a faster time scale [Fig.~\ref{relax}(b)] than the population within the Dirac cone [Fig.~\ref{relax}(c)].

Even more, the scattering-mediated electron-hole recombination in Fig.~\ref{tstages}(g) will proceed on a slower time scale than the scattering shown in Fig.~\ref{tstages}(f). Electrons within the high-density surface-projected bulk conduction band will recombine as in Fig.~\ref{tstages}(f). However, when the metastable chemical potential falls below the minimum of the bulk-conduction band, electrons and holes that are left to recombine are within the TSS [Fig.~\ref{relax}(c) and Fig.~\ref{holexc}]. This recombination will be much slower due to (i) the small density of states in the Dirac cone, (ii) the stringent selection rules for scattering involving spin, energy and momentum \cite{Sanchez-Barriga-PRB2-2017}, and, (iii) because the second electron participating in the scattering can only be one of the few within the TSS [Fig.~\ref{tstages}(g)].

Finally, let us provide an estimation of the transport times for both lateral (parallel to the sample surface) and transversal (towards the depth of the bulk) diffusion. For the lateral transport to give non-negligible effects, electrons are supposed to travel a distance comparable to the pump spot size (and not the probe size) since the excitation profile is practically constant over smaller distances. Considering that the group velocity of the electrons in the relevant energy range is never above 1 nm/fs, the time that would take an electron to cross ballistically a distance of 0.1 mm is 100 ps. Obviously, due to the presence of numerous scatterings, the real characteristic transport time is much longer than 100 ps. It is then obvious that it is impossible for the lateral transport to contribute in any way. We also stress that the presence of electric fields does not change the conclusion: electric fields cannot accelerate electrons on the surface of these materials above the aforementioned maximum group velocity. The situation is very different for transverse transport where electrons and holes have to travel a distance (worst case scenario) of few tens of nm to be outside or inside of the probed volume, requiring a time scale which is fully compatible with our present findings.

\section{IV. Summary and outlook}

In conclusion, the overall dynamics observed here reveals that whenever transport is inhibited or nearly completely quenched this creates a thermalization bottleneck for the decay of excited states even at high energies above the Fermi level. In the bulk-conducting transport regime, on the other hand, the thermalization bottleneck is completely suppressed resulting in an accelerated electron dynamics. This pinpoints the crucial role of ultrafast transport in the thermalization and relaxation dynamics of excited states in TIs. Our results thus show how tr-ARPES on TIs can be directly sensitive to ultrafast transport phenomena which are otherwise difficult to access using conventional transport methods. The present findings also imply that the decay of excited states in prototypical TIs such as Bi$_2$Te$_3$, Bi$_2$Se$_3$ or Sb$_2$Te$_3$ cannot be understood solely on the basis of elementary scattering processes such as electron-electron and electron-phonon scatterings. Therefore, it would be interesting to perform similar experiments as the ones reported here in ultrathin TI films by taking advantage of compositionally graded doping, or to explore the impact of laser-induced out-of-equilibrium transport at interfaces between TIs and trivial insulators.

In the context of applications, the present findings offer a promising route to control the relaxation time scales of light-induced spin currents and spin-polarized electrical currents on the surface of TIs, which is of relevance for spintronics. This concept could be also utilized to print circuits based on TIs that can be photoactivated paving the way for geometrical engineering of  ultrafast information transport. In addition, increasing the relaxation time scales of electron and holes within the TSS can be important in the context of energy material applications, such as solar cells. Recently, TIs have been predicted to be intrinsic solar cells reaching an efficiency of about 7\% \cite{Fregoso-JPCM-2015}. Therefore, the present results offer the possibility of exploiting ultrafast transport to tailor the capabilities of TIs as intrinsic solar cells.

\section{Acknowledgements}

Financial support from SPP 1666 of the Deutsche Forschungsgemeinschaft is gratefully acknowledged. M. B. gratefully acknowledges the Austrian Science Fund (FWF) through Lise Meitner position M1925-N28 and Nanyang Technological University, NAP-SUG for the funding of this research.

\end{document}